\documentclass[floatfix,aps,twocolumn,floatfix,10pt,superscriptaddress,prl]{revtex4-2}

\usepackage{bm}
\usepackage{graphicx}
\usepackage{subfigure}
\usepackage{wrapfig}
\usepackage{epstopdf}
\usepackage{siunitx}
\usepackage{braket}
\usepackage{tabularx}
\usepackage{blindtext}
\usepackage{amsmath}
\usepackage{float}

\usepackage{color}
\usepackage[section]{placeins}

\def\bem#1{\begin{mathletters}\label{#1}}
\def\eml{\end{mathletters}}

\def\4#1{{\boldsymbol{#1}}}
\def\8#1{{\widetilde{#1}}}

\newcommand{\simgeq}{\; \raisebox{-0.4ex}{\tiny$\stackrel
{{\textstyle>}}{\sim}$}\;}

\begin{document}

\title {Nitrogen-vacancy singlet manifold ionization energy}

\author{S. A. Wolf}
\affiliation{The Racah Institute of Physics, The Hebrew University of Jerusalem, Jerusalem 91904, Israel}
\affiliation{The Center for Nanoscience and Nanotechnology, The Hebrew University of Jerusalem, Jerusalem 91904, Israel}

\author{I. Meirzada} 
\affiliation{The Racah Institute of Physics, The Hebrew University of Jerusalem, Jerusalem 91904, Israel}

\author{G. Haim} 
\affiliation{Dept. of Applied Physics, Rachel and Selim School of Engineering, Hebrew University, Jerusalem 91904, Israel}

\author{N. Bar-Gill}
\affiliation{The Racah Institute of Physics, The Hebrew University of Jerusalem, Jerusalem 91904, Israel}
\affiliation{The Center for Nanoscience and Nanotechnology, The Hebrew University of Jerusalem, Jerusalem 91904, Israel}
\affiliation{Dept. of Applied Physics, Rachel and Selim School of Engineering, Hebrew University, Jerusalem 91904, Israel}

\begin{abstract}
The singlet states of the negatively-charged nitrogen-vacancy centers in diamond play a key role in its optical spin control and readout.
In this work, the hitherto unknown ionization energy of the singlet is measured experimentally and found to be between 1.91-2.25 eV. This is obtained by analyzing photoluminescence measurements incorporating spin control and NV
charge state differentiation, along with simulations based on the nitrogen-vacancy's master equation. This work establishes a protocol for a more accurate estimate of this ionization energy, which can possibly lead to improved read-out methods.

\end{abstract}

\maketitle

\section{Introduction}

Negatively charged Nitrogen vacancy (NV$^{-}$) centers are point defects in diamond which show promising quantum properties essential for various applications ranging from magnetic sensing to quantum computing. These applications rely on the NV center's 
exceptionally long coherence times and optical spin readout, even at room temperature. The properties of the NV center are constantly being studied to improve its coherence time and readout signal-to-noise ratio (SNR). While most of the energy levels and transition rates have been measured, some are still unknown and may have a significant impact on our understanding of the system's dynamics.

\begin{figure}[tbh]
{\includegraphics[width = 0.9 \linewidth]{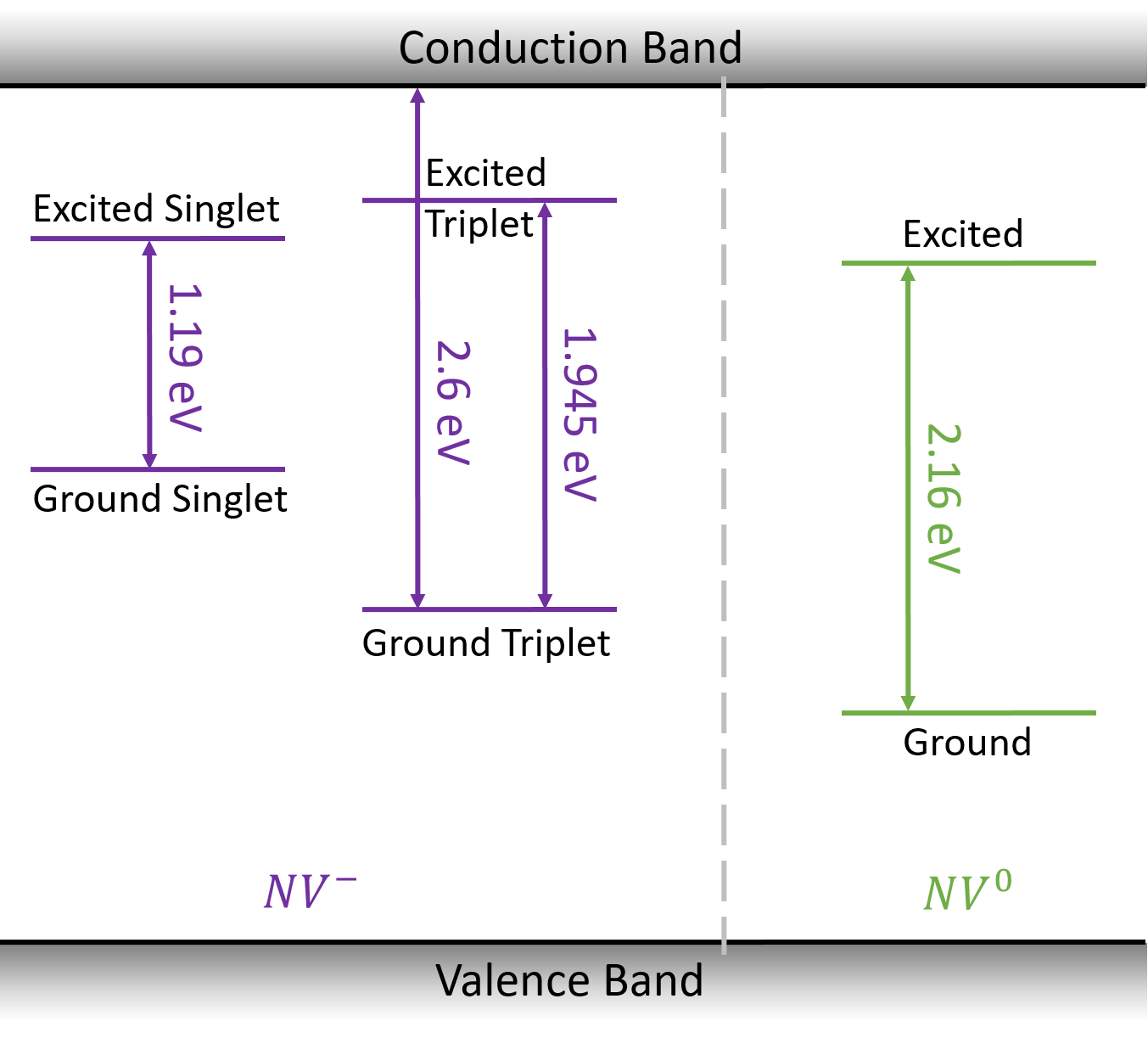}}
\caption{Known energy levels of NV centers.}
\label{fig:Energy}
\end{figure}

The NV center is a defect in diamond where two adjacent carbons are replaced by one nitrogen and a vacancy. The negatively charged NV has an additional electron which leads to two holes in the valence shell and an effective two-electron system. The ground state of the NV$^{-}$ is a spin one triplet with a 2.87 GHz zero-field-splitting between spin projections $m_s = 0$ and $m_s = \pm1$, and can be controlled using a resonant microwave (MW) drive. These spin states are located 2.6 eV below the conduction band (Fig. \ref{fig:Energy}). The electronically excited triplet state is located 1.945 eV above the ground state with a lifetime of $\sim$13 ns \cite{Davies1976,Collins1983}. The radiative transition between the two triplet states is dominated by the phonon side band (PSB), with $\sim$650-800 nm emission. The NV$^{-}$ has an additional singlet (spin zero) manifold consisting of two energy levels that are separated by 1.19 eV with an excited state lifetime of $\sim$0.1 ns \cite{Rogers_2008,Acosta_optical_2010,Kehayias_2013,Ulbricht_Excited_state_lifetime_2018}. Spin selective intersystem crossing (ISC) allows the excited triplet to decay non radiatively to the singlet excited state predominantly from the $m_s = \pm1$ spin projection. The ground singlet state, commonly referred to as a metastable state due to its long lifetime, $\sim$300 ns, decays back to the ground triplet state. Theoretical studies have predicted different values for the energy of the singlet state \cite{Drabenstedt_Singlet_1999, Gali_AbInitio_2008, Zyubin_Chemical_2009, Delaney_Singlet_2010, Toyli_Singlet_2012, Goldman_SSISC_2015, Goldman_Phonon_2015, Rogers_Singlet_2015, Thiering_Singlet_2018, Bhandari_Singlet_2021}. \emph{However, the energy gap between the singlet manifold and the conduction band has yet to be measured experimentally, and is the focus of this work}. The NV$^-$ can also be ionized optically, and convert to a neutral NV (NV$^0$) center which has an energy gap of 2.16 eV between it's ground and excited states.

\begin{figure*}[th]
\centering
    \includegraphics[width = 0.95 \textwidth, trim={0 0.5cm 0 0}]{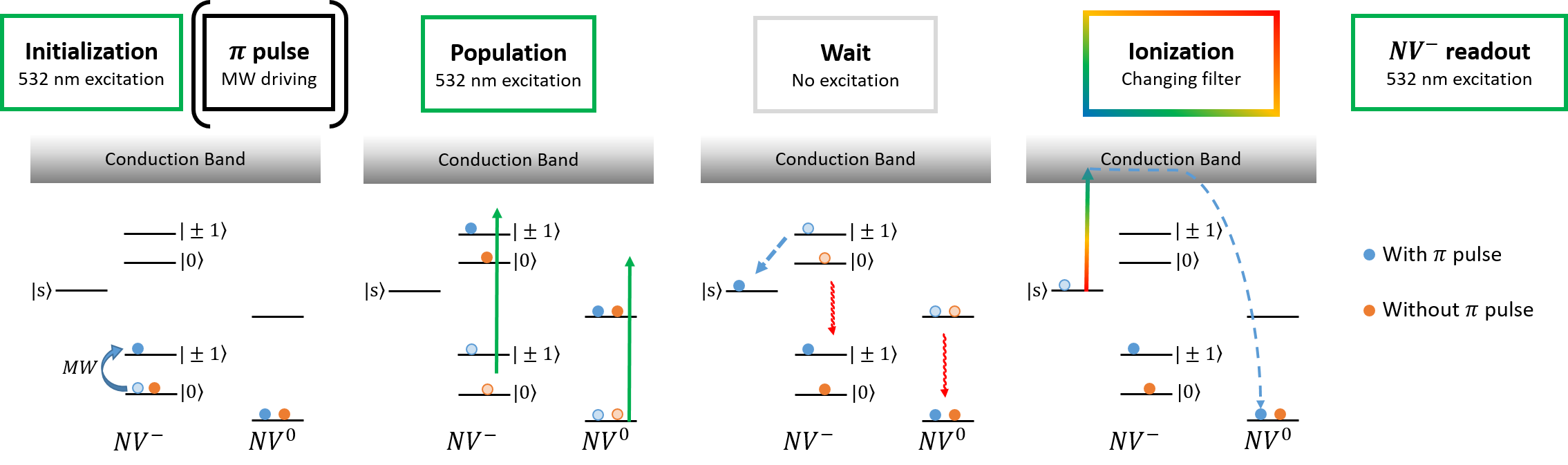}
\caption{The NV center's dynamics during the pulse sequence. Blue (orange) dots represent the population with (without) a $\pi$ pulse in the initialization step. The faded dots are a reminder of the population before the current pulse. For simplicity, the ionization pulse shows only the potential ionization from the singlet. Depending on the wavelength, additional transitions may occur during this pulse such as NV$^{-}$/NV$^{0}$ excitation and ionization/recombination from the excited states (which are not shown in this figure).}
\label{fig:Sequence}
\end{figure*}

The most commonly used readout method at room temperature is based on collecting photoluminescence (PL) from the PSB while exciting the triplet transition using the PSB (typically at 532 nm) \cite{Hopper2018}. Due to the spin selective ISC, the number of photons emitted depends on the initial spin projection of the ground triplet state. However, this readout method yields very poor SNR and requires thousands of repetitions \cite{Hopper2018}. Different readout methods have been proposed in order to increase the readout SNR \cite{Hopper2018,Meirzada_Enhanced_2019}, but those readout methods all rely on the spin-dependent transition to the metastable singlet state. Better knowledge of the energy levels and transition rates, and specifically the unknown singlet energy, will allow more accurate modeling of the NV dynamics and may lead to the development of improved readout methods. Theoretical studies have predicted various values for the energy of the singlet state, and experimental data is required in order to constrain this parameter \cite{Drabenstedt_Singlet_1999, Gali_AbInitio_2008, Zyubin_Chemical_2009, Delaney_Singlet_2010, Toyli_Singlet_2012, Goldman_SSISC_2015, Goldman_Phonon_2015, Rogers_Singlet_2015, Bhandari_Singlet_2021}.

In a recent paper \cite{Meirzada_FindingSinglet_2021} an experimental protocol for finding the ionization energy of the singlet level was proposed by the authors. The main idea behind this protocol is to maximally populate the singlet level before applying an ionization laser pulse at varying wavelengths to check for ionization (from NV$^{-}$ to NV$^{0}$). An analysis of the NV's master equation predicts significantly different results between ionization pulse wavelengths below and above the ionization energy. 
In this work we report first experimental results towards a direct measurement of the singlet ionization energy and provide experimental bounds to the singlet energy. Our measurement and analysis included important modifications on the original experiment proposal.  

The manuscript is organized as follows: we first detail the experimental protocol and pulse sequence used. We then describe the measured results, mostly of the ionization ratio normalized to different spin initializations, as a function of ionization laser power. We present the detailed measurement results, along with comparisons to simulations, in separate sections based on the ionization laser wavelength. Finally, we summarize and conclude.

\section{Pulse sequence}

The pulse sequence in the following measurements begins with an initialization to the ground state's $m_s =\pm1$ spin projection, i.e. initialization to spin projection $m_s=0$ with a long 532 nm pulse followed by a $\pi$ pulse using resonant MW. Next, a short 532 nm population pulse optimized to maximally populate the ground singlet state is applied (200 $\mu$W for 400 ns, see Appendix B for details), followed by a $\sim$30 ns delay to allow the triplet excited states to fully decay to either the singlet or ground triplet states. At this point the singlet is maximally populated and an ionization pulse is applied while the singlet is still highly populated ($\sim$ 100 ns). Finally, a 532 nm readout pulse is applied, during which the emitted NV$^-$ PL is collected using a 650 nm long-pass filter (see figure 2 in Ref. \cite{Meirzada_FindingSinglet_2021}). 

Assuming the singlet manifold lies between the ground and excited triplet states, the ionization energy of the ground singlet state is bound between  1.84 eV, corresponding to 674 nm, and 2.6 eV, corresponding to 477 nm \cite{Meirzada_FindingSinglet_2021}. In the following experiments we use a pulsed supercontinuum laser (NKT Photonics, WL-SC-400-15-PP) followed by different bandpass filters (BPF) (Blue - 500$\pm$20 nm, Green - 550$\pm$20 nm, Red - 650$\pm$20 nm, Long Red - 676$\pm$4 nm) and a near-infrared (NIR) 976 nm continuous-wave (CW) laser (BL976-PAG900) for the ionization pulse. The Long Red filter and NIR laser are not expected to ionize from the singlet ground state and were added as control measurements. The experiments were conducted on a diamond sample with a high density of NV centers. During the experiment the PL was measured as a function of the ionization pulse power. A reference measurement without a $\pi$ pulse in the first initialization step (i.e. initialized to $m_s = 0$ in the first step of the experiment) was conducted for every ionization laser power in order to help isolate the singlet contribution from the rest of the NV dynamics. The results both with and without the $\pi$ pulse are normalized by the PL measurement without an ionization pulse (ionization pulse power = 0 mW), and with the same initial state. In the following section we present the PL as a function of the ionization pulse power with different filters, as well as the ratio between the experimental results with and without the $\pi$ pulse in the initialization step (which will be referred to as PNP ratio).

\section{results}

\begin{figure}[tbh]
{\includegraphics[width = 1 \linewidth]{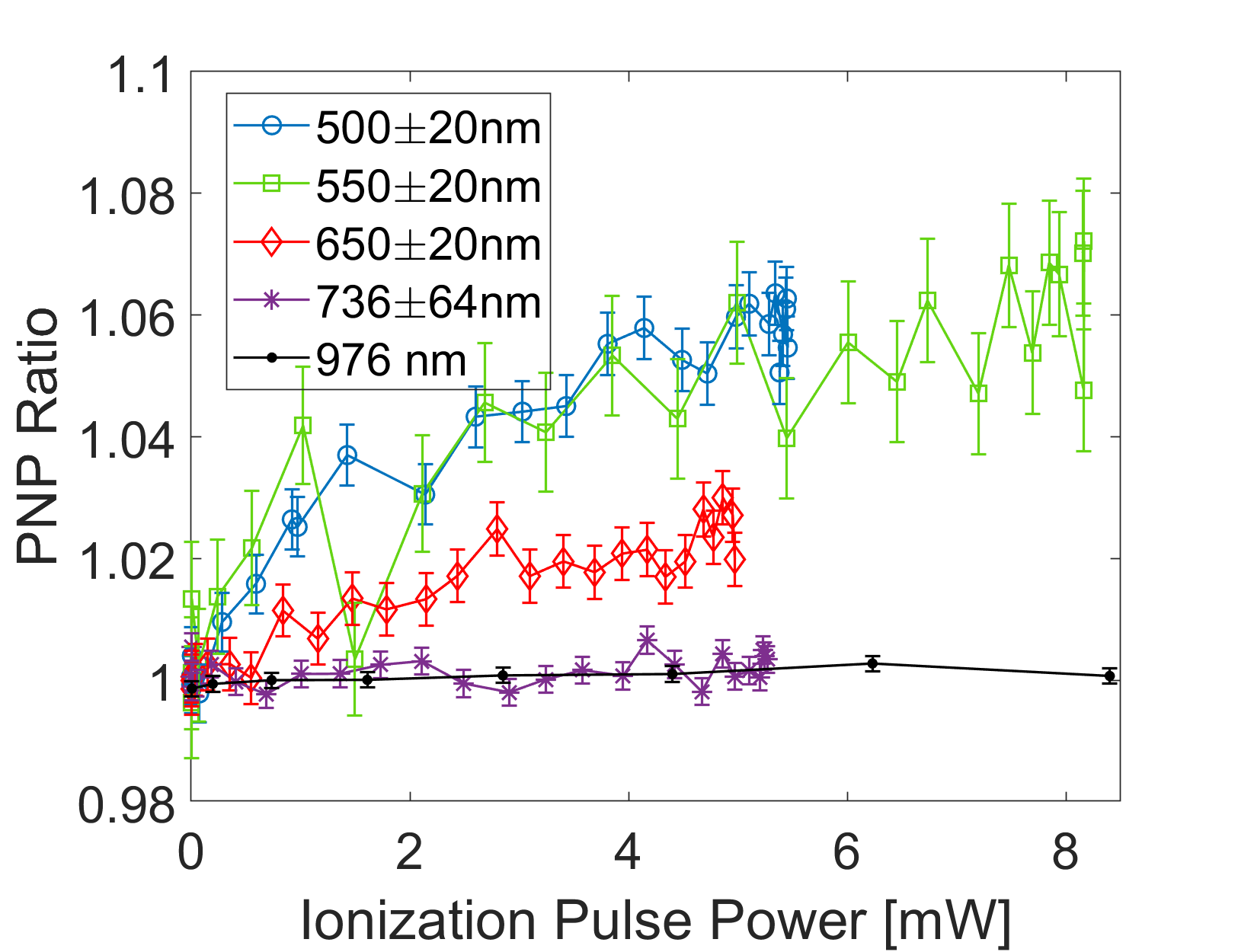}}
\caption{PNP ratio experimental results. The ratio between experimental results with and without $\pi$ pulse in the initialization step with different ionization laser filters. Blue circles shows experimental results with blue filter, green squares with green filter, red diamonds with red filter, purple stars with long red filter, and black dots with NIR laser. The solid lines were added between the experimental measurements as a visual aid.}
\label{fig:ExpPiRatio}
\end{figure}

Figure \ref{fig:ExpPiRatio} presents the PNP ratio as a function of the ionization pulse power, allowing us to visually compare the different wavelength results. This summarizes the measurements (detailed below) and provides one of the main results of this work. The results with green and blue ionization pulses show very similar behaviour, suggesting they have the same (or very similar) singlet ionization cross sections (SICS). The PNP ratio with the red filter increases less than with the blue and green filters. However, since the red filter no longer excites the NV$^0$ ground state, a more careful analysis is needed to interpret this effect. The PNP results of the two control experiments, presented in purple (long red) and black (NIR) curves, show little to no increase and are $\sim$ 1 for all laser powers, as expected (since they do no excite both NV charge states).

In the following sections we compare the experimental results with simulations, and fit the SICS in order to obtain a clear conclusion regarding the ionization energy. The simulations are based on the NV center's master equation (detailed in appendix A). Since we use a 532 nm excitation in all the experiments we start by examining the green filter.

\subsubsection{Green - 550$\pm$20 nm}

\begin{figure}[tbh]
\subfigure[]
{\includegraphics[width = 0.9 \linewidth]{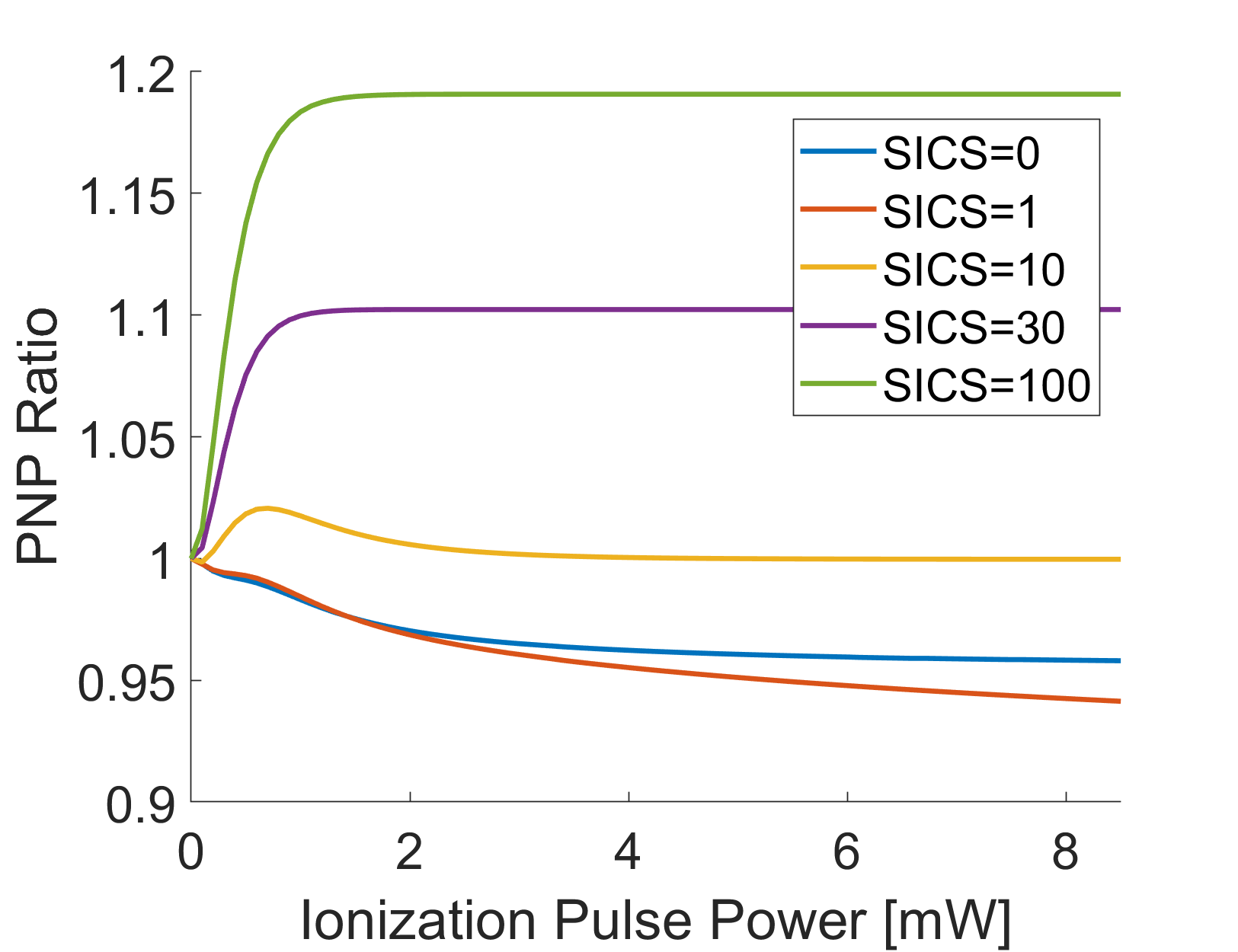}}
\subfigure[]
{\includegraphics[width = 0.9 \linewidth]{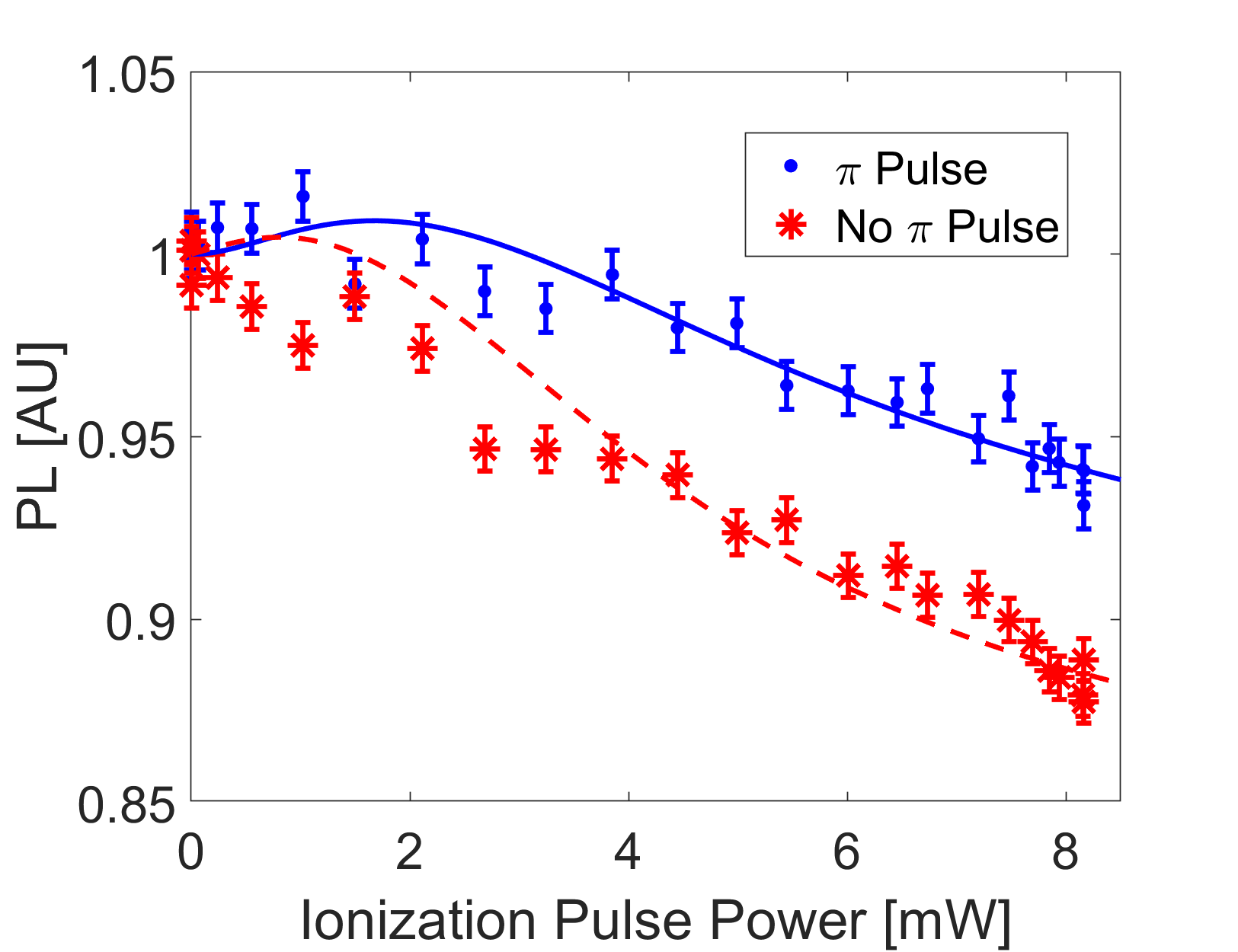}}
\caption{Simulation and experimental results with green filter. Simulations assume the green filter and the 532 nm excitation laser have the same SICS. (a) Simulation of PNP ratio with different SICS. Values of SICS in legend are in units of MHz/mW. (b) Experimental results with (blue dots) and without (red stars) $\pi$ pulse and simulations using fitted parameters for power factor and SICS (blue solid line - with $\pi$ pulse, and red dashed line - without $\pi$ pulse).}
\label{fig:Green}
\end{figure}

Figure \ref{fig:ExpPiRatio} (green line) and \ref{fig:Green}(a) show the experimental results and simulations of the PNP ratio with a green ionization pulse, respectively. We can see that if there was no ionization from the singlet (SICS = 0) with this filter the PNP ratio should decrease as a function of the ionization pulse power and be $<$ 1. The increase in the PNP ratio in the experimental results indicates that there is a non-zero ionization cross section from the singlet. We can therefore conclude that the singlet ionization energy must be $\lesssim$ 2.25 eV (corresponding to 550 nm).

We note that the PNP ratio in the simulation seems to reach a saturation value faster than in the experimental results. This could be the result of the focal spot of the ionization laser not being diffraction limited and the use of a very dense NV sample; both would lead to an effectively lower excitation power. Therefore, an ionization power scaling parameter was added to our simulations. 

In order to obtain the value of the green SICS we consider the experimental results with and without the $\pi$ pulse separately, depicted in figure \ref{fig:Green}(b). Based on this data, we find the best fitting parameters for the green filter to be SICS $= 20.0 \pm 0.5$ MHz/mW and power scaling $= 0.11 \pm 0.0015$. These simulation results are also shown in figure \ref{fig:Green}(b), as solid blue ($\pi$ pulse) and dashed red (no $\pi$ pulse) lines. In the following sections we will assume that SICS $= 20$ MHz/mW for the 532 nm excitation laser.
We note that the simulation assumes a perfect $\pi$ pulse and therefore the real SICS might be slightly higher than our extracted value. 

\subsubsection{Blue 500$\pm$20 nm}

\begin{figure}[tbh]
\subfigure[]
{\includegraphics[width = 0.9 \linewidth]{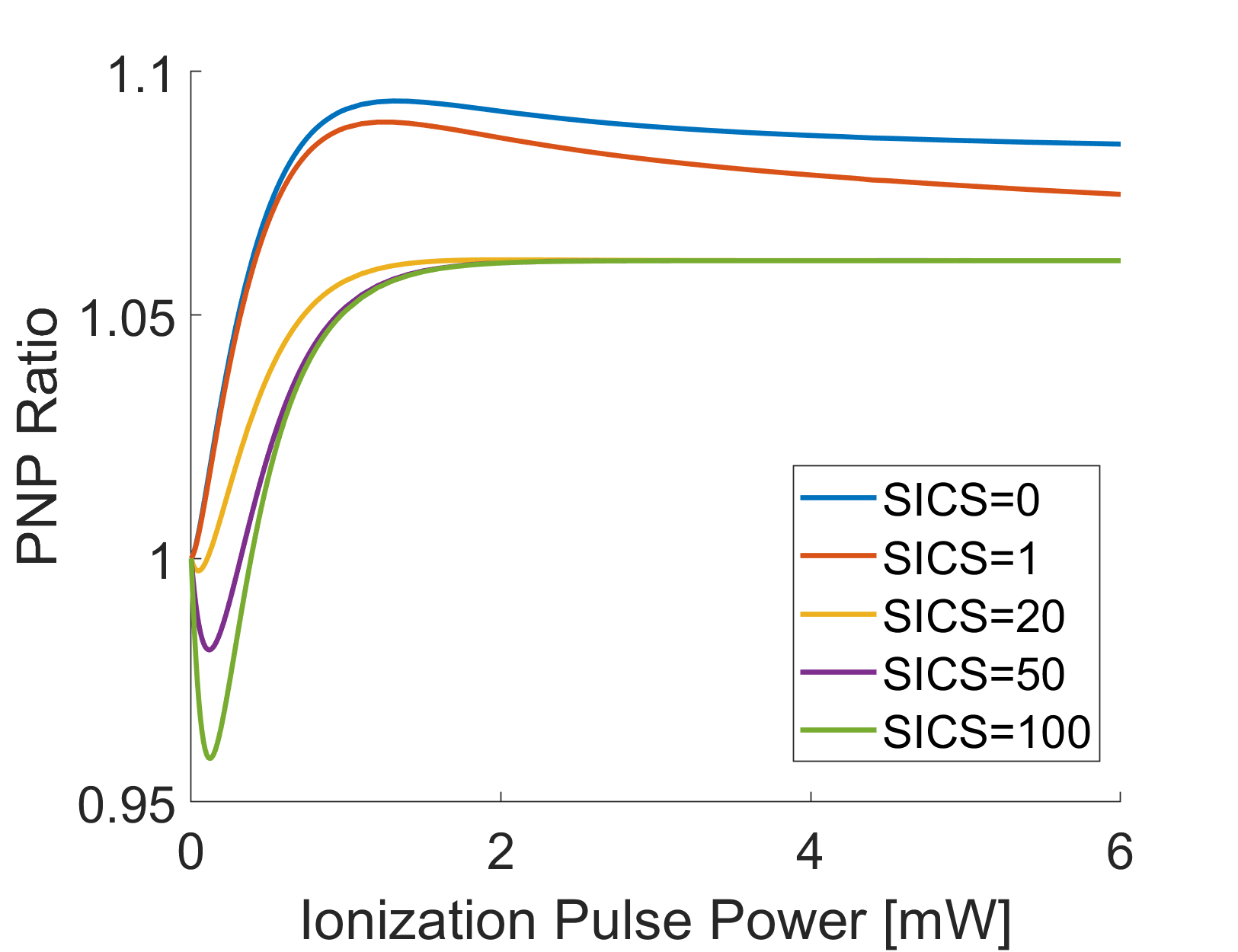}}
\subfigure[]
{\includegraphics[width = 0.9 \linewidth]{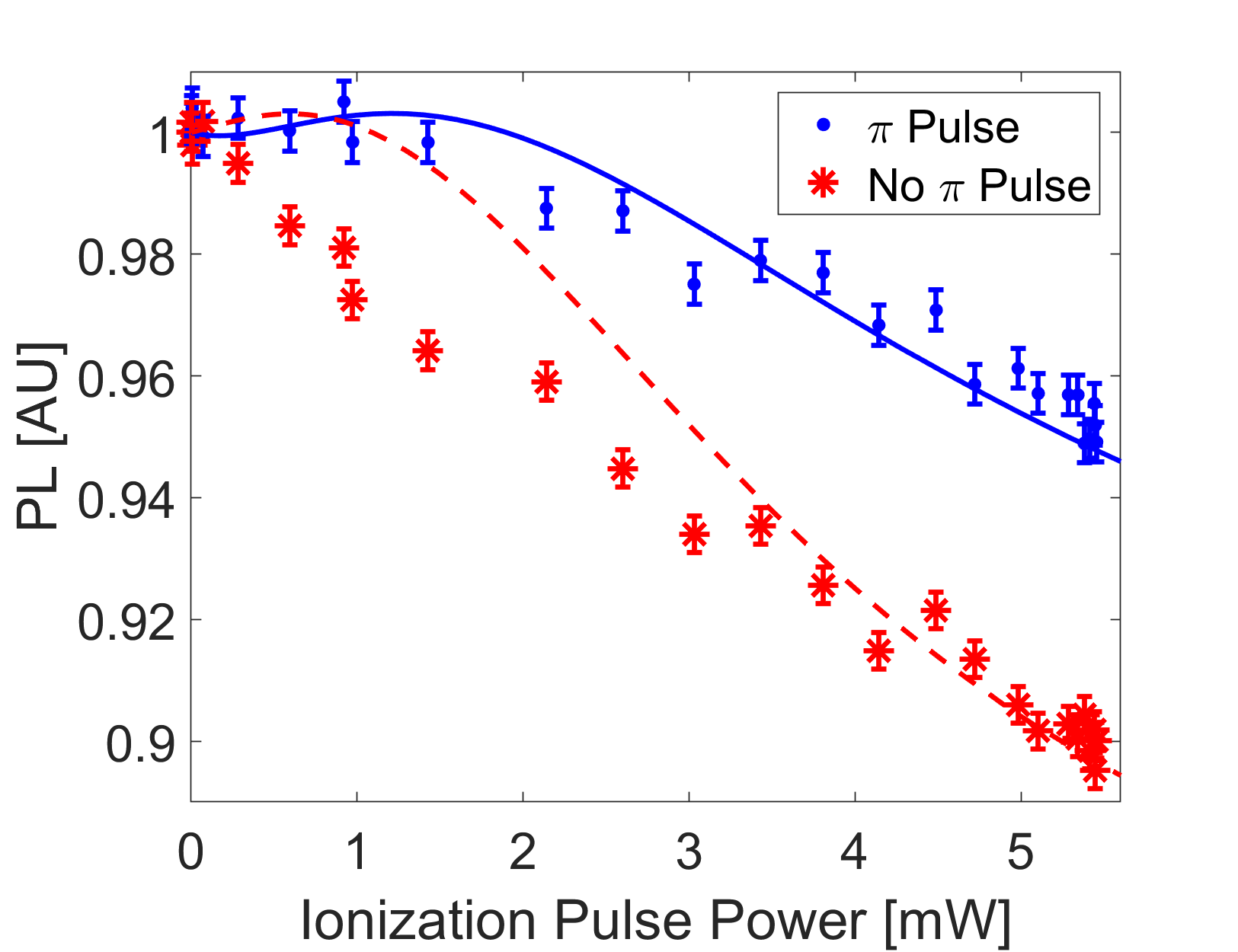}}
\caption{Simulation and experimental results with Blue filter. Simulations assume SICS = 20 MHz/mW for the 532 nm excitation laser. (a) Simulation of PNP ratio with different SICS. Values of SICS in legend are in units of MHz/mW. (b) Experimental results with (blue dots) and without (red stars) $\pi$ pulse and simulations using fitted parameters for power factor and blue SICS (blue solid line - with $\pi$ pulse, and red dashed line - without $\pi$ pulse).}
\label{fig:Blue}
\end{figure}

As discussed above, the experimental PNP ratios with blue and green ionization filters (blue and green lines in figure \ref{fig:ExpPiRatio}, respectively) almost all fall within the error bars of one another. Assuming the NV excitation rates and ionization/recombination rates from the excited states with the blue filter are the same as the green, this indicates that the blue SICS $\sim$ green SICS. This can also be seen from figure \ref{fig:Blue}(a), in which we show simulations of the PNP ratio with a blue ionization pulse and different blue SICS. For small SICS values the simulations show a maximum that does not appear in the data, while for large enough SICS values the PNP ratio decreases below one which is also inconsistent with the data.

Assuming that the NV excitation and ionization/recombination rates with the blue filter are the same as the green, we fit the simulations with and without $\pi$ pulse to the data separately (Fig. \ref{fig:Blue}(b)). As with the green ionization pulse, we have two fitting parameters, the blue SICS and power scaling. The fit gives an optimum at blue SICS $= 22.8 \pm 0.8$ MHz/mW and power scaling $= 0.129 \pm 0.004$. As expected, the blue and green SICS are equal within error bars. The power scaling parameter is also similar in both cases.

\subsubsection{Red - 650$\pm$20 nm}

\begin{figure}[tbh]
\subfigure[]
{\includegraphics[width = 0.48 \linewidth]{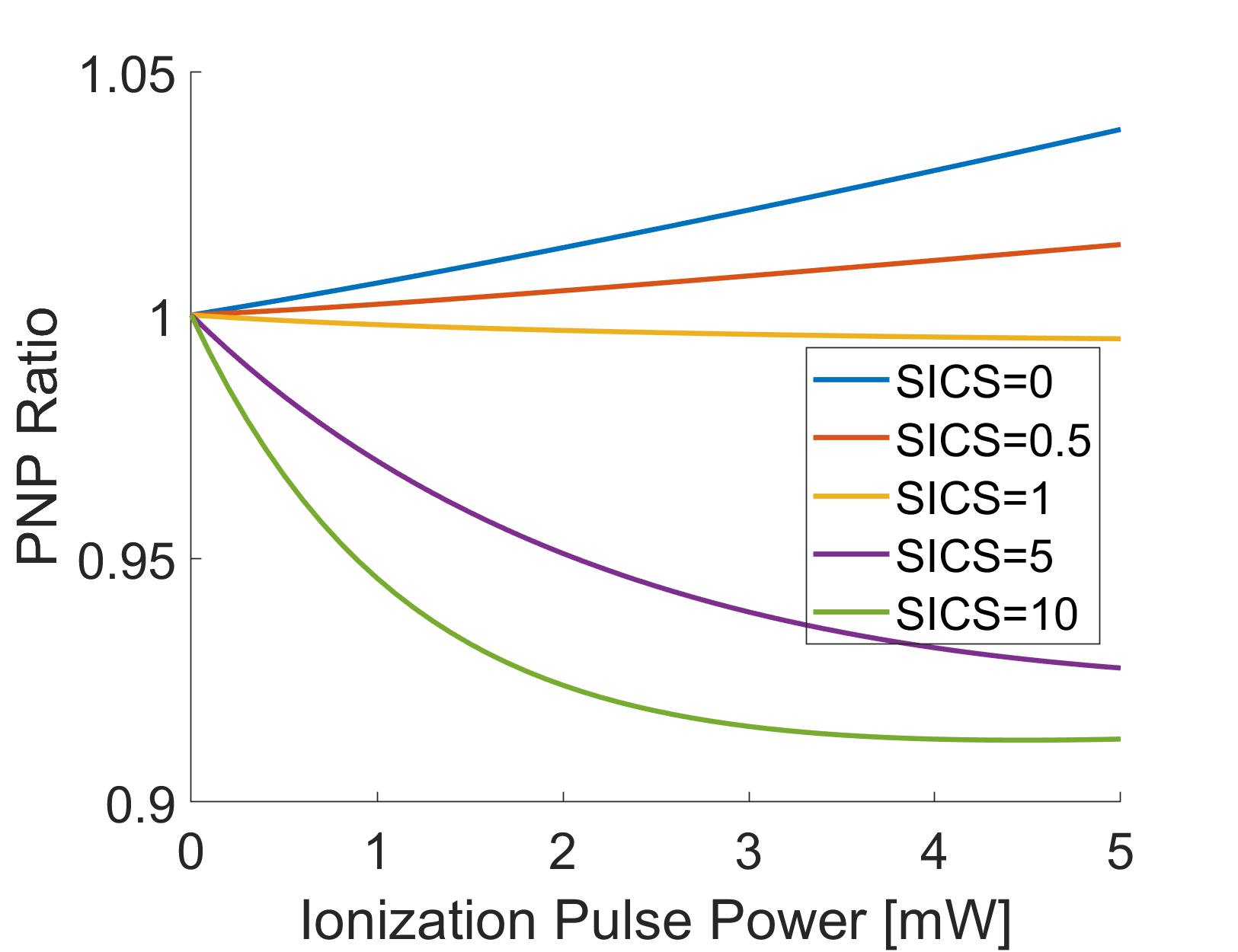}}
\subfigure[]
{\includegraphics[width = 0.48 \linewidth]{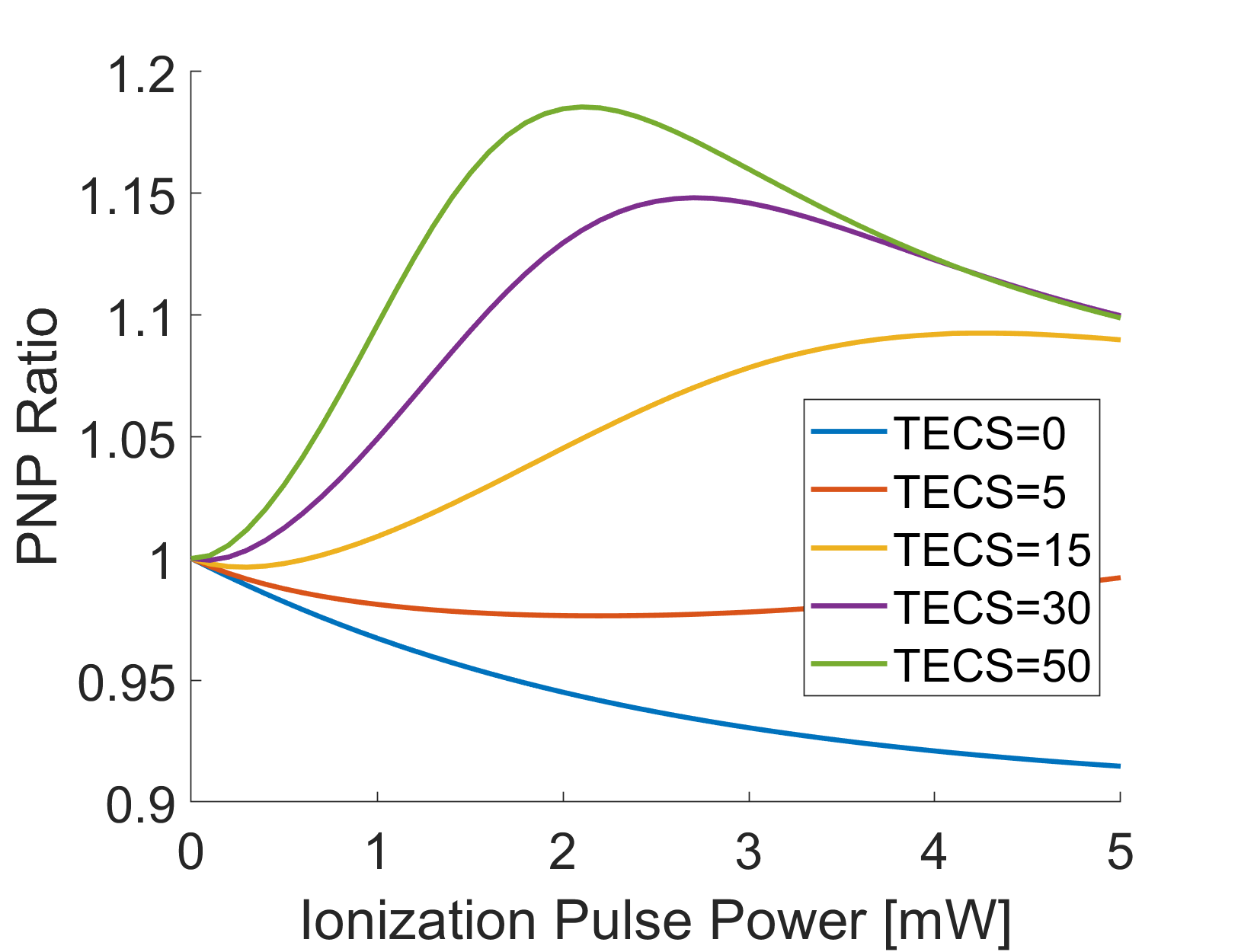}}
\subfigure[]
{\includegraphics[width = 0.9 \linewidth]{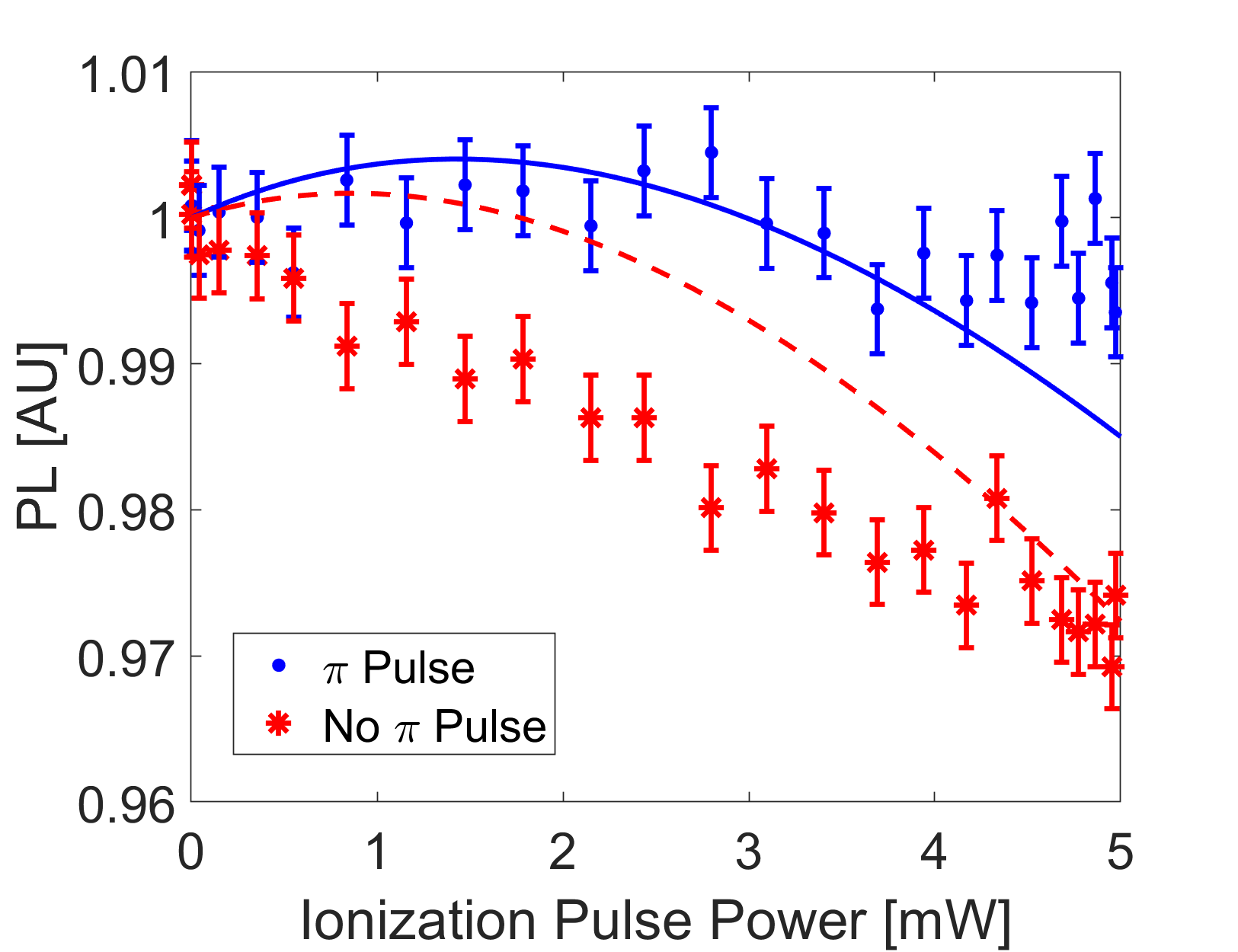}}
\caption{Red filter - simulations and experimental results. (a) PNP ratio simulations assuming TECS = 1 MHz/mW and different values for red SICS. Values of SICS in legend are in units of MHz/mW. (b) PNP ratio simulations assuming red SICS = 5 and different values of TECS. Values of TECS in legend are in units of MHz/mW. (c) Experimental results with and without $\pi$ pulse in the initialization step (blue dots and red stars, respectively) and simulation results using fitted parameters for red SICS, red TECS, red triplet ionization cross section and red recombination from NV$^0$ (with $\pi$ - blue solid line, and without $\pi$ - red dashed line).}
\label{fig:Red}
\end{figure}

The experimental results with the red filter are shown in Figures \ref{fig:ExpPiRatio} (red line) and \ref{fig:Red}(c). As with the shorter wavelengths, the ratio here is also greater than 1 and increases with the ionization pulse power, but this increase is less significant. Note that this filter can excite the NV$^-$ triplet ground state to the triplet excited state (ZPL$^{-}$ = 637 nm), while it \emph{does not} excite the NV$^0$ (ZPL$^0$ = 575 nm). The effective triplet excitation cross section (TECS) with this filter was not directly measured, and is therefore another unknown parameter in our simulations (bound by the known 532 nm TECS). 

Figures \ref{fig:Red}(a) and (b) present the simulation results with a red ionization pulse and show a strong dependence of the results on the red SICS and TECS, respectively. These simulations assume that the ionization and recombination rates from the excited states are the same as those with 532 nm laser \cite{meirzada_negative_2017}. Assuming the red TECS = 1 MHz/mW (Figure \ref{fig:Red}(a)), one may conclude that the red SICS must be $< 0.5$ MHz/mW, to obtain the experimentally observed increase in the PNP ratio as a function of ionization pulse power as observed experimentally. However, one could assign the red SICS a value of 5 MHz/mW (Figure \ref{fig:Red}(b)), and by changing the red TECS, a PNP ratio $>$ 1 can be achieved. Thus the red SICS cannot be uniquely determined based on the PNP ratio results alone, unlike for previous (bluer) wavelengths.

In order to determine the red SICS, we fit our simulation to the experimental results using four fitting parameters: red SICS and TECS, and red ionization and recombination rates from the excited states (NV$^-$ excited triplet and NV$^0$ excited state, respectively). We assume the ionization power scaling with this filter is the same as with the green filter. We find that the optimal red SICS is $0.006^{+0.044}_{-0.006}$ MHz/mW, meaning \emph{there is no (or negligible) ionization from the ground singlet state with this filter}. The optimal red TECS = $26 \pm 5$ MHz/mW, the ionization rate from the excited triplet is $8.2^{+1.0}_{-0.6}$ MHz/mW and the recombination rate from the NV$^0$ excited state is $0.008^{+3.744}_{-0.008}$ MHz/mW. The red TECS is indeed smaller than the green TECS and consistent with \cite{Subedi2019_TECS}. The red ionization and recombination rates from this fit are within error between those of 532 nm and 1064 nm \cite{meirzada_negative_2017,Beha_Excitation_2012}. \emph{The results from this section indicate that the ionization energy is $\simgeq$1.91 eV (corresponding to 650 nm)}.

\subsubsection{Control Experiments} 

Two control experiments where performed with a long red filter (672-800 nm BPF), and a NIR 976 nm CW laser, such that for both cases the SICS should be zero. Figure \ref{fig:ExpPiRatio} (purple and black lines) shows the PNP ratio with the long red filter and NIR laser, respectively. Both these ratios are $\sim$1 with very small dependence on laser power. Since both these wavelengths do not excite the NV center (either charge state), these results indicate that the SICS $\simeq$ 0, as expected. 

\begin{figure}[tbh]
\subfigure[]
{\includegraphics[width = 0.9 \linewidth]{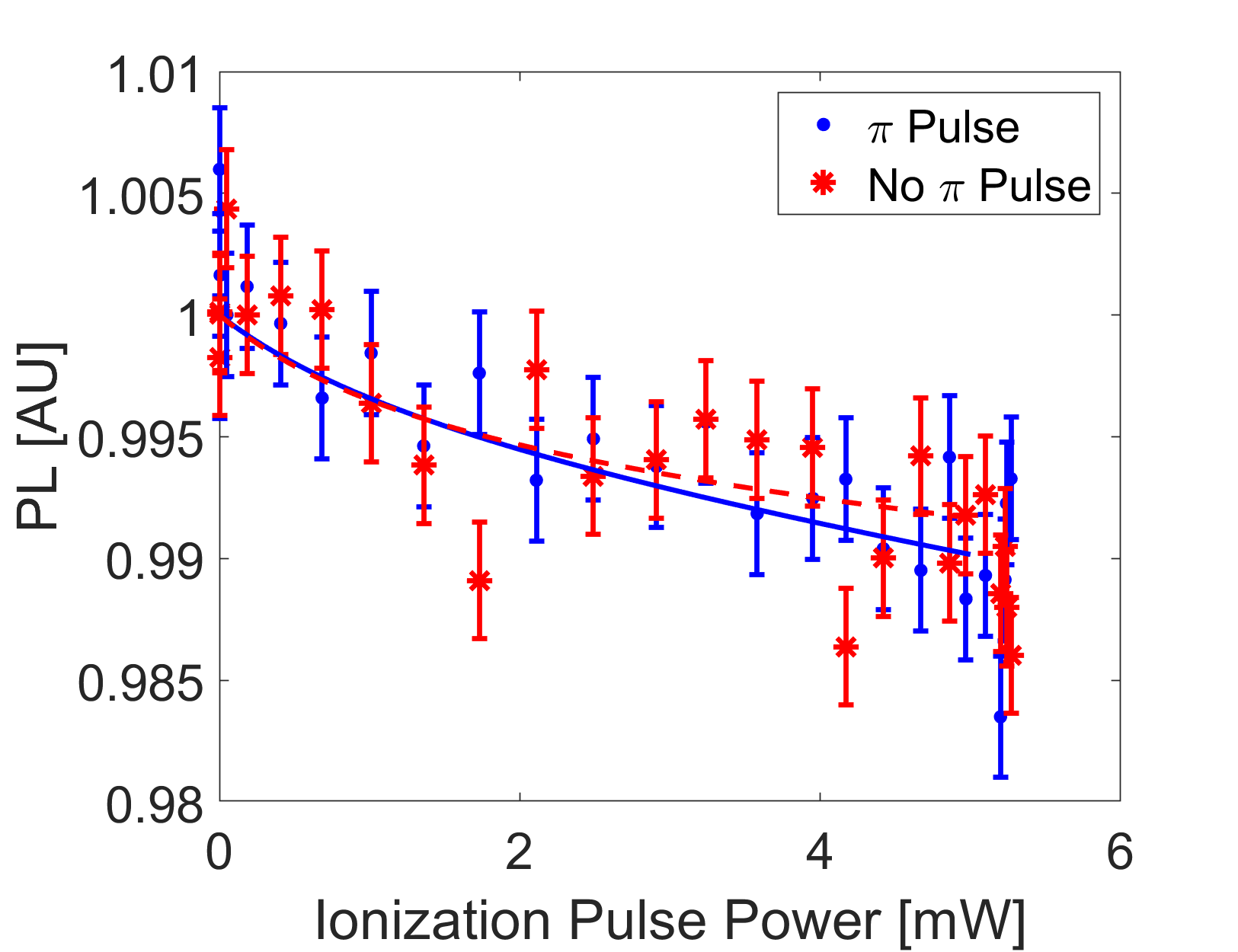}}
\caption{Long red filter - experimental results and simulations. Experimental results with and without $\pi$ pulse in the initialization step (blue dots and red stars, respectively) and simulation results using fitted parameters for long red SICS, triplet ionization cross section and recombination from NV$^0$ (with $\pi$ - blue solid line, and without $\pi$ - red dashed line).}
\label{fig:LongRed}
\end{figure}

The experimental results for the long red filter with and without a $\pi$ pulse are shown in Figure \ref{fig:LongRed}, with fitted simulations. If there was no population in the excited states during the ionization pulse, we would have expected to see no effect from the ionization pulse, i.e. PL = 1 independent of ionization pulse power. The small decay observed in the experiments is a result of the little population remaining in the excited states before the ionization pulse is applied. The optimal parameters found to fit this data are long red SICS = 0.08 $\pm$ 0.04 MHz/mW, excited triplet ionization cross section = 28 $\pm$ 10 MHz/mW and recombination cross section = $0^{+2}_{-0}$ MHz/mW. For this filter the ionization power scaling was assumed to be one. Changing the power scaling has little to no effect on the optimal long red SICS and mostly changes the optimal triplet ionization cross section. As expected, the optimal SICS is very close to zero.

\begin{figure}[tbh]
{\includegraphics[width = 0.9 \linewidth]{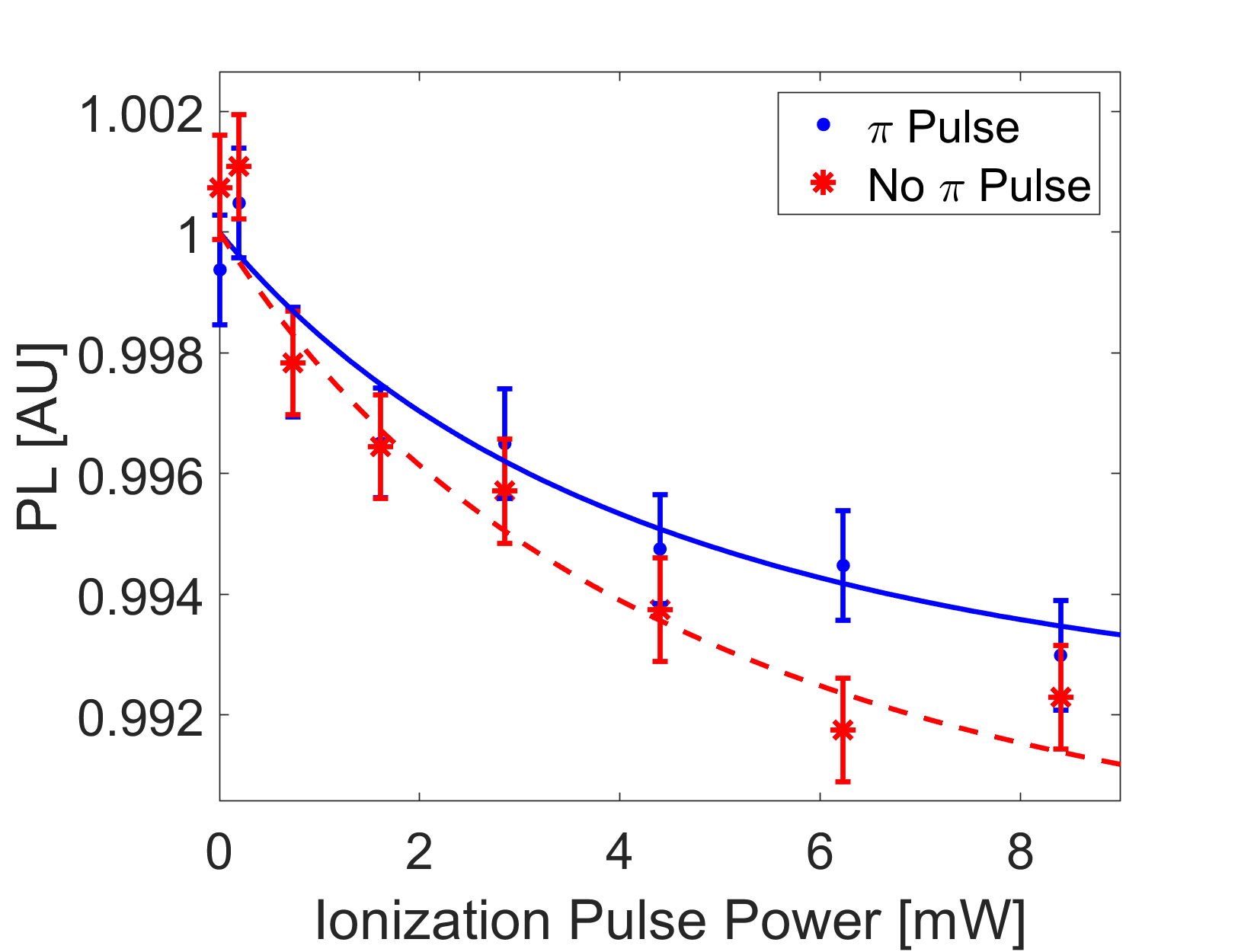}}
\caption{NIR experimental results and simulations. Experimental results with and without $\pi$ pulse in the initialization step (blue dots and red stars, respectively) and simulation results with fitted parameters with (blue solid) and without (red dashed) $\pi$ pulse in the initialization step.}
\label{fig:NIR}
\end{figure}

Figure \ref{fig:NIR} depicts the experimental results and simulations with fitted parameters for the NIR laser. The simulations include excitation of the singlet ground state to the singlet excited state with NIR \cite{Kehayias_2013}, however, do not include an ionization rate from the excited singlet state with NIR since this effect should be negligible given the experiment's ionization powers \cite{hopper_near-infrared-assisted_2016}. The fitting shows optimal NIR SICS = 0.002$^{+0.05}_{-0.002}$ MHz/mW, NIR ionization from NV$^-$ excited triplet = 13$^{+10}_{-4}$ MHz/mW, and NIR recombination from NV$^0$ excited state = 0.2$^{+1.3}_{-0.2}$ MHz/mW. The optimal NIR SICS is consistent with zero, as expected.

\section{Summary}

In this work we develop a tailored protocol to spectroscopically identify the NV$^{-}$ singlet energy levels, and implmenet it by measuring the NV photodynamics as a function of various ionization wavelengths. Our results impose bounds on these energy levels, placing the ground singlet level between 1.91 - 2.25 eV below the conduction band.

We have followed the protocol suggested in \cite{Meirzada_FindingSinglet_2021} and incorporated several required modifications. Our results are in agreement with the theoretical predictions of \cite{Goldman_SSISC_2015, Goldman_Phonon_2015,Thiering_Singlet_2018}, and significantly improve our ability to pinpoint the singlet energy level. Additional experiments in the remaining energy range and with narrower filters (or a tunable laser) could impose an even tighter bound on the ionization energy range of the singlet.

\section{Acknowledgements}

We would like to thank Prof. Yossi Paltiel and Dr. Yuval Kolodny for their help with the supercontinuum laser. We thank Yoav Ninio, Ty Zabelotsky, Pavel Penshin and Amir Chen for helping build the experimental setup. We also thank Dima Budker and Fedor Jelezko for fruitful discussions. The work of S.A.W. was supported by the Dalia and Dan Maydan fellowship and the Levitan fellowship.

N.B. acknowledges support from the European Union’s Horizon 2020 research and innovation program under grant agreements No. 101070546 (MUQUABIS) and No. 828946 (PATHOS), and has been supported in part by the Ministry of Science and Technology, Israel, the innovation authority (project \#70033), and the ISF (grants \# 1380/21 and 3597/21).

\section{Appendix A - Master equation and rates}

\begin{figure}[tbh]
{\includegraphics[width = 0.9 \linewidth]{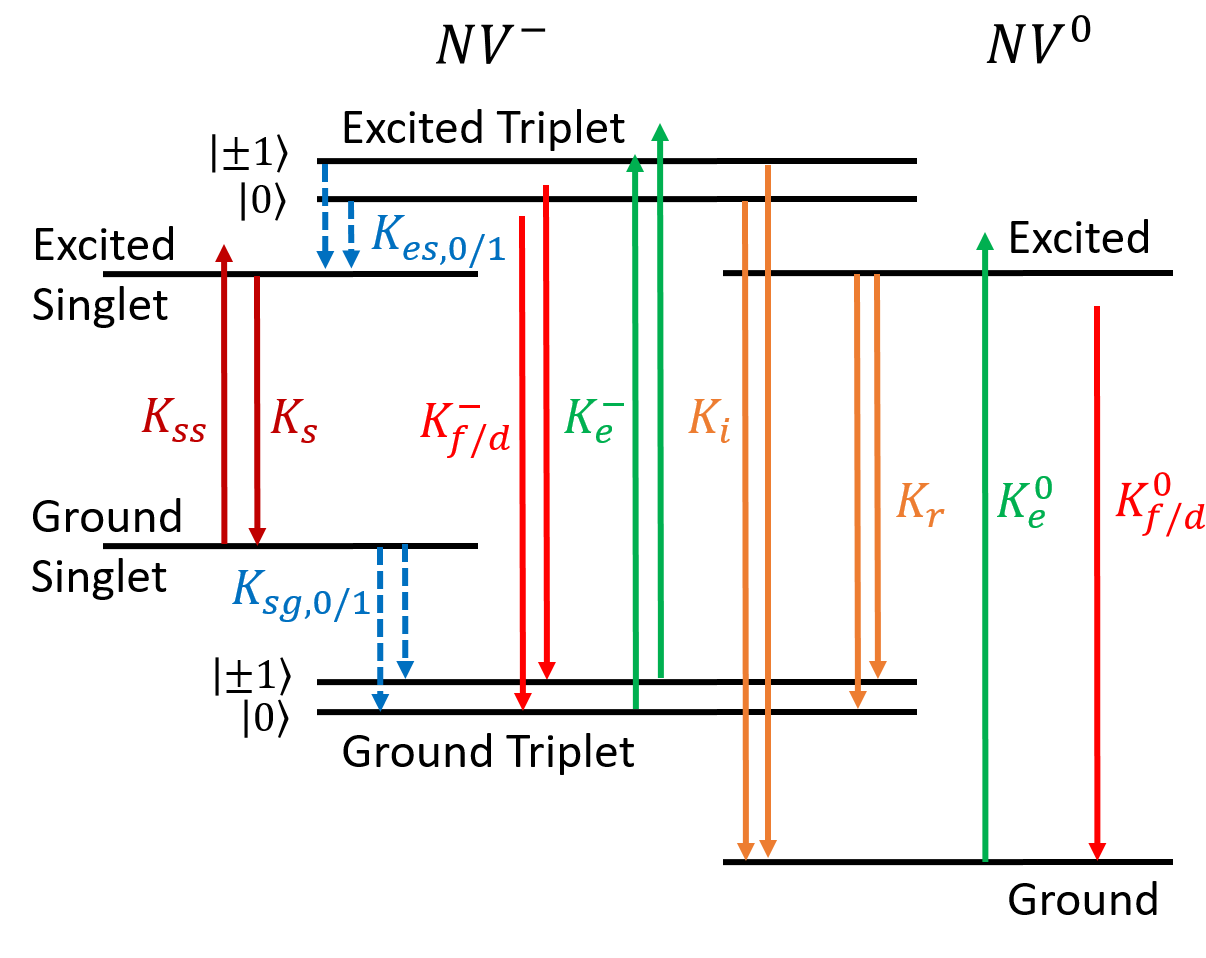}}
\caption{NV center energy levels diagram and transitions.} 
\label{fig:energyleveldynamics}
\end{figure}

The NV center 8-level model as was simulated in this paper is described in Fig. \ref{fig:energyleveldynamics}. 

The NV dynamics are governed by the following master equation:
\\
\\
\noindent ${ \dot{P}^-_{g,0} = -K^-_eP^-_{g,0}+(K^-_f+K^-_d)P^-_{e,0}+K_{sg,0}P^-_{g,s}+K_rP^0_e }$

\noindent ${ \dot{P}^-_{g,1} = -K^-_eP^-_{g,1}+(K^-_f+K^-_d)P^-_{e,1}+K_{sg,1}P^-_{g,s}+K_rP^0_e}$

\noindent ${ \dot{P}^-_{e,0} = -(K^-_f+K^-_d+K_{es,0}+K_i)P^-_{e,0}+K^-_eP^-_{g,0} }$

\noindent ${ \dot{P}^-_{e,1} = -(K^-_f+K^-_d+K_{es,0}+K_i)P^-_{e,1}+K^-_eP^-_{g,1} }$

\noindent ${{\dot{P}^-_{e,s} = -K_{s}{P^-_{e,s}}+K_{ss}P^-_{g,s}+K_{es,0}(P^-_{e,0}+P^-_{e,1}) }}$

\noindent ${ \dot{P}^-_{g,s} = -(K_{sg,0}+K_{sg,1}-K_{ss})P^-_{g,s}+K_{s}{P^-_{e,s}} }$

\noindent ${ \dot{P}^0_g = -K^0_eP^0_g+(K^0_f+K^0_d)P^0_e+K_i(P^-_{e,0}+P^-_{e,1}) }$

\noindent ${ \dot{P}^0_e = -(K^0_f+K^0_d+2K_r)P^0_e+K^0_eP^0_g }$
\\

The populations of the NV states are denoted by $P^{0/-}_n$, where the upper index indicates the charge state of the NV. The letter "g" ("e") in the lower index stand for the ground (excited) state, singlet states are denoted by "s" in the lower index, and triplet spin projection is denoted by "0" ("1") for $m_s = 0$ ($m_s = \pm1$). The rates are denoted by the letter K and are specified in figure \ref{fig:energyleveldynamics}. The simulations rely on previously published charge dynamics \cite{meirzada_negative_2017}, recently measured singlet decay rate \cite{Ulbricht_Excited_state_lifetime_2018}, and internal NV$^-$ rates from \cite{Robledo_Dynamics_2011}, and stimulated emission cross sections from \cite{Fraczek_depletion_2017}, as detailed in tables \ref{table:DecayRates} and \ref{table:ExcitationRates}.

\begin{table}[tbh]
  \begin{tabular}{c|c}
      Transition & Rate \\
      \hline
      $K^-_f$ & 77 MHz \\
      ${K^0_f}$  & 53 MHz \\
      ${K_{es,0}}$ & 0 MHz \\
      ${K_{es,1}}$ & 30 MHz \\
      ${K_{s}}$ & 10000 MHz \\
      ${K_{sg,0}}$ & 3.3 MHz \\
      ${K_{sg,1}}$ & 0 MHz \\
      \hline
  \end{tabular}
\caption{The internal NV$^-$ and NV$^0$ decay rates that were used to simulate the NV center's dynamics \cite{Robledo_Dynamics_2011,meirzada_negative_2017,Ulbricht_Excited_state_lifetime_2018}}
\label{table:DecayRates}
\end{table}

\begin{table}[tbh]
  \begin{tabular} {c|c|c|c|c} 
      Transition & Green/Blue & Red & Long Red & NIR \\
      \hline
      $K^-_e$ & 135 & fitted & 0 & 0 \\
      ${K^0_e}$ & 243 & 0 & 0 & 0 \\
      ${K_{ss}}$ & 0 & 0 & 0 & 0.92 \\
      ${K_{i}}$ & 43 & 43 (fitted) & fitted & fitted \\
      ${K_{r}}$ & 17.75 & 17.75 (fitted) & fitted & fitted \\
      ${K^-_{d}}$ & 0 & 10 & 30 & 0 \\
      ${K^0_{d}}$ & 0 & 18 & 12 & 0 \\
      \hline
  \end{tabular}
\caption{The internal NV$^-$ and NV$^0$ excitation rates, and ionization and recombination rates used in the simulations for the different wavelengths \cite{Robledo_Dynamics_2011,meirzada_negative_2017,Fraczek_depletion_2017}. The rates in this table are in units of MHz/mW, and were multiplied by the excitation power in the master equation.}
\label{table:ExcitationRates}
\end{table}

\section{Appendix B - Population Pulse Calibration}

We tried to maximize the population of the singlet during the ionization pulse in order to get a maximal contrast. The singlet population was measured by excitation of the singlet transition (976 nm, off resonance excitation) and collection of the singlet's PSB (1000 nm long pass filter), which will be referred to as singlet readout.

\begin{figure}[tbh]
{\includegraphics[width = 0.9 \linewidth]{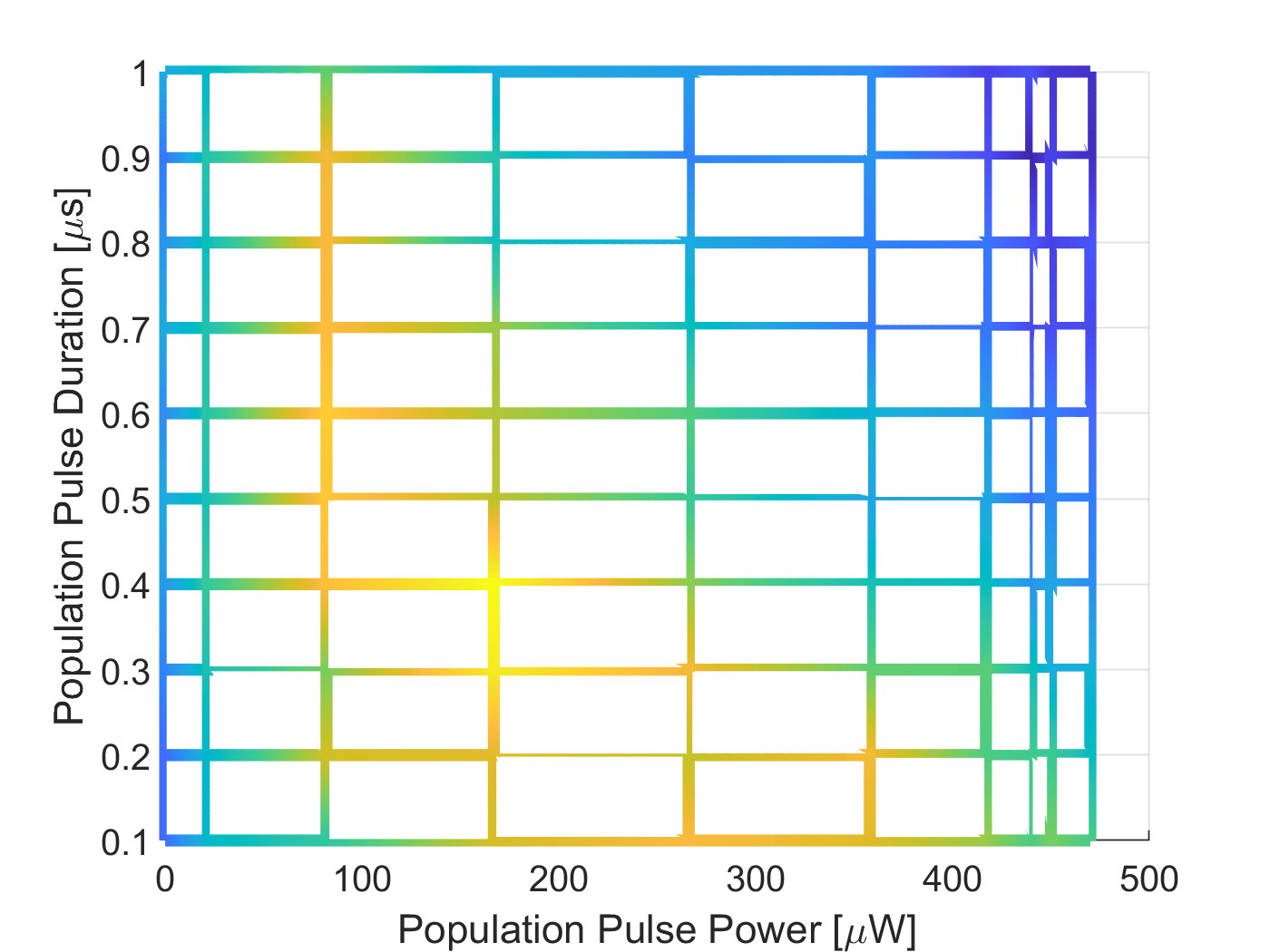}}
\caption{Optimization of population pulse parameters. PL from the singlet transition (proportional to singlet population during ionization pulse) as a function of pulse duration and power.}
\label{fig:PopCalib}
\end{figure}

The power and duration of the population pulse were optimized together as they depend on each other. The optimization was done using the same pulse sequence as the singlet ionization experiment described in the main text, except that the readout was done during the ionization pulse and measured the singlet population (singlet readout). The Population pulse duration and power were changed throughout the experiment to measure their effect on the singlet population during the ionization pulse. As can be seen in Fig. \ref{fig:PopCalib}, the optimal population pulse duration is 400 $\mu s$ and with 200 $\mu W$ excitation laser power. 

\FloatBarrier

\bibliography{refs1.bib}

\end{document}